\documentstyle[12pt]{article}
\def\be{\begin{equation}}
\def\ee{\end{equation}}
\begin{document}
\hfill IFT-P.054/92

\hfill hep-ph/9212269

\hfill December 1992
\vskip 1.5cm
{\Large \center {\bf Comment on ``Chiral Dilepton Model and The Flavor
Question''\\}}
\vskip 1cm
Recently in Ref.~\cite{pf} a model was considered based
on the
$$SU(3)_c \otimes SU(3)_L \otimes U(1)_X$$
 gauge symmetry.
This model is identical to a model previously proposed in
Ref.~\cite{ppp,pp1} and has precisely the same gauge group and
representation content.Here we just list some features associated
with this model and address the reader to Refs.~\cite{ppp,pp1,pp2}
for details. In particular we correct some incorrect results and
misconceptions of Ref.~\cite{pf}.
The model has extra gauge bosons with single and double charge
$Y^-,Y^{--}$, called dileptons in Ref.~\cite{pf}
($V^-,U^{--}$ in Refs.~\cite{ppp,pp1}) and also a
new neutral gauge boson $Z'^0$.
Both kinds of vector bosons gain mass mainly from the vacuum
expectation value (VEV), $w$ of the Higgs triplet $({\bf3},+1)$ which
precipitates the first symmetry breaking. Ref.~\cite{pf}, using LEP
measurement of the quantity $1-M^2_W/M^2_Z$, obtained the lower limit
on the new gauge bosons:
\be
M_{Z'}\geq 300\, \mbox{GeV}\quad M_Y\geq 230\, \mbox{GeV}.
\label{lower}
\ee
The $Z'^0$ boson couples to flavor changing neutral currents which
give a contribution to the $K^0-\bar K^0$ mass difference and this
imposes the stronger limit on its mass given in Ref.~\cite{pp1}:
\be
M_{Z'}>40\,\mbox{TeV}.
\label{lb1}
\ee
This implies that $w$ is of the order of $8$ TeV and consequently
\be
M_{Y^-,Y^{--}}>4\, \mbox{TeV}.
\label{lb2}
\ee
On the other hand, taking into account the three triplets as well as the
sextet needed in the model in order to give mass to all
the fermions, we found that~\cite{pp2}
\be
\frac{M^2_Z}{M^2_W}=\frac{1+4t^2}{1+3t^2},
\label{1}
\ee
where $t=g^2_X/g^2$. However, in
order to be consistent with experimental data Eq.(\ref{1}) must be
numerically equal to $1/\cos^2_W$, where $\theta_W$ is the weak
mixing angle in the standard electroweak model. This implies
\be
t^2=\frac{\sin^2\theta_W}{1-4\sin^2\theta_W}.
\label{2}
\ee
Ref~\cite{pf} assumed a normalization for the $U(1)_X$
and $T_3$ generators, prompted by a potential unification theory,
which imposes the condition
\be
g^2_X/g^2=4/5.
\label{3}
\ee
However, from Eq.~(\ref{2}),
and using $\sin^2\theta_W\simeq0.23$ we obtain $g^2_X/g^2\simeq2.87$
which is in conflict with (\ref{3}). Then, only the lower bounds,
Eq.~(\ref{lb1} and (\ref{lb2}) on the masses of the new gauge bosons
arise.

The parameter $\xi=(M^2_Z/M^2_{Z'})_0$, where the subscript denotes
``lowest order'', was used in Ref.~\cite{pf} in order to study the
corrections to the standard model predictions. It was found that
\be
\xi=\frac{g^2(3+8t^2)}{4(1+2t^2)}\frac{v^2}{w^2},
\label{xi}
\ee
where $v$ and $U$ are, respectively, the vacuum expectation value of
the triplet with $X=0$ and $X=+1$. Since
$w>8\,\mbox{TeV}$ and as we must consider $v$ is of the order of the
Fermi scale we see that the $\xi$-parameter is too small to be
measured in currently energies and the new bosons will be detected
only indirectly at new generations high-energy colliders.

I thank O.F. Hern\'andez and R. Foot for useful discussions. This  work
was  partially supported by  the
Con\-se\-lho Na\-cio\-nal de De\-sen\-vol\-vi\-men\-to Cien\-t\'\i \-fi\-co e
Tec\-no\-l\'o\-gi\-co (CNPq).

\noindent{\bf V. Pleitez}\\
Instituto de F\'\i sica Te\'orica\\
Universidade Estadual Paulista, Rua Pamplona, 145\\
01405-900--S\~ao Paulo, SP, Brazil.\\

\end{document}